\documentclass[12pt]{article}

\usepackage{mathrsfs}
\usepackage{fullpage}
\usepackage{amsfonts}
\usepackage{graphicx}
\usepackage{mathrsfs}
\usepackage{amsmath}
\usepackage{algorithm}
\usepackage{algpseudocode}
\usepackage{amssymb}
\usepackage{float}
\usepackage{subfig}
\usepackage{rotating,color}
\usepackage{xcolor,natbib,epstopdf}
\usepackage{appendix}
	\renewcommand\appendix{\par
	\setcounter{section}{0}
	\setcounter{subsection}{0}
	\setcounter{lemma}{0}
	\setcounter{equation}{0}
	\renewcommand{\thelemma}{\arabic{lemma}}
	\gdef\thesection{Appendix \Alph{section}}
	\gdef\thesubsection{\Alph{section}.\arabic{subsection}}}
\renewcommand\arraystretch{0.8}

\newtheorem{theorem}{ \noindent T{\footnotesize HEOREM}}[section]

\def\bm {\boldsymbol}

\def\bms{{\bf \Sigma}}
\def\S{{\bf S}}

\def\O{{\bf \Omega}}
\def\I{{\bf I}}
\def\U{\bm U}
\def\bth{\bm \theta}
\def\bT{{\bf \Theta}}

\def\X{\bm X}

\def\G{{\bf \Gamma}}
\def\Z{\bm Z}
\def\L{{\bf \Lambda}}
\def\B{{\bf B}}
\def\A{{\bf A}}
\def\M{{\bf M}}

\def\bp{{\bf \Psi}}
\def\var{{\rm Var}}
\def\cov{{\rm Cov}}
\def\tr{{\rm tr}}

\def\cd{\mathop{\rightarrow}\limits^{d}}

\def\vec{\operatorname{vec}}
\def\vech{\operatorname{vech}}

\title
{\bf Adaptive Sphericity Tests for High Dimensional Data}
\author{Ping Zhao$^{1}$, Wenwan Yang$^{2}$, Long Feng$^{1}$ and Zhaojun Wang$^{1}$\\
$^{1}$School of Statistics and Data Science, KLMDASR, LEBPS, and LPMC,\\ Nankai University\\
$^{2}$School of Mathematics and Statistics, Guangdong University of Technology
}
\date{}

\begin{document}
\maketitle

\begin{abstract}
In this paper, we investigate sphericity testing in high-dimensional settings, where existing methods primarily rely on sum-type test procedures that often underperform under sparse alternatives. To address this limitation, we propose two max-type test procedures utilizing the sample covariance matrix and the sample spatial-sign covariance matrix, respectively. Furthermore, we introduce two Cauchy combination test procedures that integrate both sum-type and max-type tests, demonstrating their superiority across a wide range of sparsity levels in the alternative hypothesis. Our simulation studies corroborate these findings, highlighting the enhanced performance of our proposed methodologies in high-dimensional sphericity testing.

\end{abstract}

\section{Introduction}

Sphericity assumptions are crucial in numerous statistical problems, spanning a wide range of applications such as geostatistics, paleomagnetic studies, animal navigation, astronomy, wind direction data analysis, and microarray analysis. The importance of testing for sphericity in these fields is well-documented; for more detailed information, please refer to the works of \cite{Tyler1987}, \cite{Baringhaus1991}, \cite{Mardia2000}, \cite{Marden2002}, \cite{Sirkia2009}, \cite{Chen2010b}, \cite{Yang2021}.

When the dimension is fixed, likelihood ratio tests are commonly employed in practice \citep{Anderson2003}. \cite{Sugiura1968} modified the biased likelihood ratio criteria. \cite{John1971,John1972} developed test statistics based on the trace of the square of the sample covariance matrix under the assumption of normality. For elliptical distributions, \cite{Muirhead1980} proposed a modification of John's test statistic, while \cite{Hallin2006} and \cite{Sirkia2009} constructed nonparametric tests using multivariate sign- or rank-based covariance matrices.

High-dimensional data are increasingly prevalent across various fields, including finance, genomics, and image processing. In these contexts, the dimension of observations is often assumed to approach infinity, rendering many traditional statistical methods ineffective. For high-dimensional sphericity testing, \cite{Bai2009}, \cite{Wang2013}, and \cite{Yang2021} have addressed the likelihood ratio tests in scenarios where the ratio
$p/n\to c\in (0,1)$. Meanwhile, \cite{Ledoit2002} evaluated John's test when
$c$ is a finite constant. \cite{Wang2013}, \cite{Li2016} further proposed corrections to both the likelihood ratio test and John's test for sphericity without relying on the normality assumption. Additionally, \cite{Chen2010b} and \cite{Ahmad2016} introduced alternative estimators for the trace of the population covariance matrix based on U-statistics. Building on this work, \cite{Xu2024} made adjustments to the aforementioned U-statistic to accommodate elliptical distributions.

All the aforementioned tests are sum-type test statistics, which aggregate information from each variable. As is widely known, these types of tests perform well under dense alternative hypotheses, where the number of nonzero signals is comparable to the total number of variables. However, for sparse alternative hypotheses, where only a few variables have nonzero signals, sum-type tests do not perform very well. In such cases, max-type tests exhibit better performance, as evidenced in \cite{Cai2014} and \cite{feng2022high}. Therefore, we have developed a max-type test procedure based on the sample covariance matrix (abbreviated as NM hereafter), assuming an independent component model. We demonstrate its advantages under sparse alternatives.

In practice, it is often challenging to determine whether the alternative hypothesis is dense or sparse. Consequently, numerous adaptive test procedures have been developed by combining sum-type tests and max-type tests, as seen in the works of \cite{xu2016adaptive}, \cite{feng2022a}, \cite{wang2023}, and \cite{Feng2024}, among others. Inspired by these ideas, we have demonstrated that the sum-type test statistic introduced by \cite{Wang2013} is asymptotically independent of the newly proposed max-type test statistics under both the null hypothesis and specific alternative hypotheses. Building upon this finding, we have proposed a Cauchy combination test procedure (abbreviated as CN hereafter) that integrates the $p$-values from these two types of tests. Both theoretical analysis and simulation studies have shown that our proposed adaptive test exhibits strong performance across a wide range of alternative hypotheses.

All the aforementioned test procedures are constructed based on the sample covariance matrix. Although some literature has extended these test statistics to elliptical distributions, their performance remains unsatisfactory. In fact, under the assumption of elliptical distributions, spatial-sign and spatial-rank methods exhibit greater efficiency, as demonstrated by \cite{oja2010}. Specifically, for the sphericity test problem, \cite{Marden2002}, \cite{Hallin2006}, and \cite{Sirkia2009} proposed spatial-sign or spatial-rank tests. However, as the dimension increases, the estimation of the spatial median introduces a non-negligible bias term in these tests. To address this issue, \cite{Zou2014} developed a bias-corrected test statistic that is robust in high-dimensional settings. Nonetheless, their test is a sum-type test procedure, which may not perform well under sparse alternatives.

Therefore, we also propose a max-type test procedure based on the sample spatial-sign covariance matrix (abbreviated as SM hereafter). We demonstrate its asymptotic independence from the test statistic proposed by \cite{Zou2014}. For alternative hypotheses with general sparsity levels, we further provide a Cauchy combination test procedure (abbreviated as CS hereafter). Both theoretical results and simulation studies indicate that the newly proposed SM test is powerful under sparse conditions, and the CS test exhibits good performance across a wide range of sparsity levels for alternative hypotheses under the elliptical distribution assumption.

This paper is structured as follows: Section 2 presents the proposed test procedures and their corresponding theoretical results. Section 3 outlines the simulation studies conducted to evaluate the performance of the proposed methods. Finally, we conclude our findings in Section 4. The appendix provides all the detailed proofs for the theoretical results.

\section{Test Procedures}
Let $\X_1,\cdots,\X_n$ be $p$-dimensional random vectors with location parameter $\bth \in \mathbb{R}^p$ and covariance matrix $\bms\in \mathbb{R}^{p\times p}$. We say a $p$-dimensional random vector is spherical if the distribution of $\X-\bth$ is invariant under orthogonal transformations, which is equivalent to test the following hypothesis:
\begin{align}\label{h0}
H_0: \bms=\sigma^2\I_p~~versus~~ H_1:\bms\not=\sigma^2\I_p
\end{align}
Under the normality assumption, John (1971,1972) proposed the following test statistic:
$$
Q_{\mathrm{J}}=\frac{n p^2}{2} \operatorname{tr}\left\{\frac{\S}{\operatorname{tr}(\S)}-\frac{1}{p} \I_p\right\}^2
$$
where $\S$ is the sample covariance matrix. Ledoit and Wolf (2002) extend this method to high dimensional case. If $\X_i\sim N(\bth,\sigma^2\I_p)$, they prove that
\begin{align*}
2p^{-1}Q_J-p\cd N(1,4).
\end{align*}
For the independent component model $\X_i=\bth+\bms^{1/2}\Z_{i}$ and $Z_i=(Z_{i1},\cdots,Z_{ip})^T$, $z_{ij}$ are i.i.d and $E(z_{ij})=0,\var(z_{ij})=1, E(z_{ij}^4)=3+\beta<\infty$, Wang and Yao (2013) show that
\begin{align}\label{cj}
2p^{-1}Q_J-p\cd N(\beta+1,4).
\end{align}
The nuisance parameter $\beta$ is estimated by $\hat{\beta}=\frac{1}{np}\sum_{j=1}^p\sum_{i=1}^n\hat{\sigma}_j^{-4}(X_{ij}-\bar{X}_i)^4-3$ and $\hat{\sigma}_j^2$ is the sample variance of $X_{1j},\cdots,X_{nj}$. Here we define the corresponding test statistic (abbreviated as NS hereafter):
\begin{align}
T_{NS}=\frac{(n-1)Q_J}{np}-\frac{p+\hat{\beta}+1}{2}.
\end{align}

However, $Q_J$ do not perform very well for heavy-tailed distributions. For elliptical distributions, the spatial sign always has good performance in a wide range of application. Hallin and Paindaveine (2006), Sirkia et al. (2009) mimicked John's test statistic and proposed the following test statistic:
$$
Q_{\mathrm{S}}=p \operatorname{tr}\left\{\operatorname{tr}\left(\O\right)^{-1} \O-p^{-1} \I_p\right\}^2=p \operatorname{tr}\left(\O-p^{-1} \I_p\right)^2
$$
where $\O=\frac{1}{n}\sum_{i=1}^n \U_i\U_i^T$, $\U_i=U(\X_i-\bth)$ and $U(\bm x)=\bm x/\|\bm x\|I(\bm x\not=\bm 0)$. When $p$ is fixed, under the null hypothesis one has
$$
n(p+2) Q_{\mathrm{S}} / 2 \cd \chi_{(p+2)(p-1) / 2}^2
$$
as $n \rightarrow \infty$. In practice, the location parameter is always unknown and replaced by its estimator--spatial median $\hat \bth$:
\begin{align*}
\hat\bth=\arg\min_{\bth \in \mathbb{R}^p} \sum_{i=1}^n \|\X_i-\bth\|.
\end{align*}
When the dimension $p$ is fixed, replacing $\bth$ with $\hat{\bth}$ does not effect the asymptotic property of $Q_S$. However, when the dimension gets larger, there would be a non-negligible bias term in $Q_S$. Zou et al. (2014) proposed a bias-corrected sign-based procedure (abbreviated as SS hereafter):
\begin{align*}
T_{SS}=(\tilde Q_S-p\hat{\delta})/\sigma_S
\end{align*}
where
\begin{align*}
\tilde{Q}_S&=\frac{p}{n(n-1)} \sum_{i \neq j}\left(\hat{\U}_i^{\mathrm{T}} \hat{\U}_j\right)^2-1, \hat{\U}_i=U(\X_i-\hat{\bth}),\\
\hat{\delta}&=\frac{1}{n^2}\left(2-2\hat{c}_2\hat{c}_1^{-2}+\hat{c}_2^2\hat{c}_1^{-4}\right)
+\frac{1}{n^3}\left(8\hat{c}_2\hat{c}_1^{-2}-6\hat{c}_2^2\hat{c}_1^{-4}+2\hat{c}_2\hat{c}_3\hat{c}_1^{-5}-2\hat{c}_3\hat{c}_1^{-3}\right),\\
\hat{c}_k&=\frac{1}{n}\sum_{i=1}^n \|\X_i-\hat{\bth}\|^{-k}, \sigma_S^2=4(p-1)/(n(n-1)(p+2)).
\end{align*}
They show that under the null hypothesis and elliptical distribution assumption, $T_{SS}\cd N(0,1)$ as $n,p\to \infty$.

All the aforementioned test procedures are of the sum-type, which exhibit strong performance under dense alternatives—situations where numerous elements of $\bp={p}\bm{\Sigma}/\text{tr}(\bm{\Sigma}) - \mathbf{I}_p$ are nonzero. However, they may lack power when dealing with sparse alternatives, where only a few elements of ${p}\bm{\Sigma}/\text{tr}(\bm{\Sigma}) - \mathbf{I}_p$ are nonzero. It is widely recognized that max-type test procedures are highly effective in such sparse scenarios. Numerous studies have explored high-dimensional testing problems using max-type tests, as evidenced in works like \cite{Cai2014} and \cite{feng2022high}. In this paper, we introduce a novel max-type test procedure specifically designed for the independent component model. Additionally, we propose a max-type test based on the spatial sign for testing sphericity for elliptical distribution.

\subsection{Max-type test procedures}
Based on the sample covariance matrix, we proposed the following max-type test statistic (abbreviated as NM hereafter):
\begin{align}\label{nm}
T_{NM}=&\max\left\{\max_{1\le k\le p}\frac{n(\hat{\sigma}_k^2-p^{-1}\tr(\S))^2}{\hat{\kappa}_k},\max_{1\le i<j\le p}\frac{n\hat\sigma_{ij}^2}{\hat{\sigma}_i^2\hat{\sigma}_j^2}\right\}\nonumber\\
&-2\log(p(p+1)/2)+\log\log(p(p+1)/2)
\end{align}
where $\hat{\kappa}_k=\frac{1}{n}\sum_{i=1}^n(X_{ki}-\bar{X}_k)^4-\hat{\sigma}_k^4$, $\bar{x}_{k}=\frac{1}{n}\sum_{i=1}^n X_{ki}$, $\hat{\sigma}_i^2=\frac{1}{n}\sum_{i=1}^n(X_{ki}-\bar{X}_k)^2$ and $\hat{\sigma}_{ij}$ is the $(i,j)-$th component of the sample covariance matrix $\S$.

Note that the sphericity test differs from the test for mutual independence between random vectors \citep{schott2005,feng2022a,wang2024rank}. Under the assumption of normality, testing for mutual independence is equivalent to testing the correlation matrix $\mathbf{R} = \mathbf{D}^{-1/2}\bms\mathbf{D}^{-1/2}$, where $\mathbf{D}$ is the diagonal matrix of $\bms$:
\begin{align}\label{h01}
H_0: \mathbf{R} = \mathbf{I}_p \quad \text{versus} \quad H_1: \mathbf{R} \neq \mathbf{I}_p.
\end{align}
For the testing problem (\ref{h01}), \cite{jiang2004} proposed the following max-type test statistic:
\begin{align*}
T_{JM} = \max_{1 \le i < j \le p} \frac{n\hat{\sigma}_{ij}^2}{\hat{\sigma}_i^2\hat{\sigma}_j^2} - 2\log\left(\frac{p(p-1)}{2}\right) + \log\log\left(\frac{p(p-1)}{2}\right)
\end{align*}
and showed that the limiting null distribution of $T_{JM}$ is the Gumbel distribution. For the sphericity test, we also need to test whether the variances of each variable are equal. Therefore, we consider the maximum of $\frac{n(\hat{\sigma}_k^2 - p^{-1}\text{tr}(\mathbf{S}))^2}{\hat{\kappa}_k}$, which evaluates the difference between the variance of each variable and the common variance $\sigma^2$. The following theorem also establishes that the limiting null distribution of our proposed test statistic $T_{NM}$ follows the Gumbel distribution.

\begin{theorem}\label{th1}
For the independent component model $\X_i=\bth+\bms^{1/2}\Z_{i}$ and $Z_i=(Z_{i1},\cdots,Z_{ip})^T$, $z_{ij}$ are i.i.d and $E(z_{ij})=0,\var(z_{ij})=1, E(z_{ij}^4)=3+\beta<\infty, E(e^{z_{ij}t})\le e^{Kt^2}$ for some positive constant $\epsilon>0$, $K>0$, under the null hypothesis (\ref{h0}) and $\frac{\log^5(p^2n)}{n}\to 0$, we have
\begin{align*}
P\left(T_{NM}\le x\right)\to G(x)\doteq \exp\left(-\frac{1}{\sqrt{\pi}}e^{-x/2}\right).
\end{align*}
\end{theorem}

Based on Theorem \ref{th1}, we reject the null hypothesis at a significance level $\alpha$ if $T_{NM} \geq q_\alpha$, where $q_\alpha = -\log \pi - 2\log\log (1-\alpha)^{-1}$ represents the $1-\alpha$ quantile of the Gumbel distribution. Furthermore, it can be readily demonstrated that when $\|\bp\|_\infty > C\sqrt{\log p/n}$ for a sufficiently large constant $C$, the probability $P(T_{NM} > q_\alpha)$ approaches 1 as $n$ and $p$ tend to infinity. This establishes the consistency of the NM test. As evidenced in \cite{Cai2014} and \cite{feng2022high}, the lower bound $\sqrt{\log p/n}$ is optimal and cannot be further improved, thereby highlighting the optimality of our proposed max-type test--NM.

The NM test mentioned above is built upon the sample covariance matrix, which may not perform well in the presence of heavy-tailed distributions or outliers. The independent component model typically does not encompass many heavy-tailed distributions, such as the multivariate t-distribution or multivariate mixture normal distribution, both of which belong to the well-known family of elliptical distributions. Therefore, we will propose a max-type sphericity test statistic specifically tailored for elliptical distributions.

We say random vector $\X$ follows elliptical distribution $E_p(\bth,\bT)$ if the density function of $\X$ is $|\bT|^{-1/2}g(\|\bT^{-1/2}(\X-\bth)\|)$ where $\bT\in \mathbb{R}^{p\times p}$ is a scatter matrix, denoted as $\X \sim E_p(\bth,\bT)$. The covariance matrix $\bms=\cov(\X)$ is proportional to the scatter matrix $\bT$. Let $\L=p{\bf\Theta}/\tr({\bf\Theta})=p\bms/\tr(\bms)$ be the sharp matrix. So the hypothesis test problem (\ref{h0}) is equivalent to
\begin{align}\label{h02}
H_0: \L=\I_p~~versus~~ H_1:\L\not=\I_p
\end{align}
As shown in \cite{oja2010} and \cite{Zou2014}, the spatial-sign based methods are very efficient for testing sphericity. Based on the sample spatial-sign covariance matrix $\hat{\O}=\frac{1}{n}\sum_{i=1}^n U(\X_i-\hat{\bth})U(\X_i-\hat{\bth})^T\doteq (\hat \psi_{ij})_{1\le i,j\le p}$, we propose the following max-type test statistic (abbreviated as SM hereafter):
\begin{align}
T_{SM}=&\max\left\{\max_{1\le i\le p}{np(p+2)\frac{(\hat\psi_{ii}-p^{-1})^2}{2(1-p^{-1})}},\max_{1\le i<j\le p}np(p+2)\hat\psi_{ij}^2\right\} \nonumber\\
&-2\log(p(p+1)/2)+\log\log(p(p+1)/2)
\end{align}
Unlike the test statistic (\ref{nm}), we do not need to estimate the variance of $\hat \psi_{ij}$ because $\var(\hat\psi_{ii})=2(1-p^{-1})/(p(p+2))$ and $\var(\hat\psi_{ij})=1/(p(p+2))$ for $i\not=j$ under the null hypothesis.

\begin{theorem}\label{th2}
If $\X_i$ follows an elliptical distribution, under the null hypothesis (\ref{h0}) and $\frac{\log^5(p^2n)}{n}\to 0$, we have
\begin{align*}
P\left(T_{SM}\le x\right)\to G(x).
\end{align*}
\end{theorem}

Theorem \ref{th2} demonstrates that the proposed SM test has a limiting null distribution that follows the Gumbel distribution. Consequently, we reject the null hypothesis at a significance level $\alpha$ if $T_{SM} \geq q_\alpha$, where $q_\alpha = -\log \pi - 2\log\log (1-\alpha)^{-1}$. Analogously, it can also be shown that $P(T_{SM} > q_\alpha) \to 1$ when $\|\bp\|_\infty > C\sqrt{\log p/n}$ for a sufficiently large constant $C$. This confirms that the SM test is optimal for sparse alternatives as well.

\subsection{Adaptive test procedures}
In practice, we often do not know whether the alternative is dense or sparse. Therefore, relying solely on either sum-type tests or max-type tests is not sufficient. Many researchers have dedicated their efforts to combining these two types of tests to develop a new adaptive testing procedure, as evidenced in literature such as \cite{xu2016adaptive}, \cite{feng2022a}, \cite{Feng2024}, and \cite{wang2024rank}. These studies have demonstrated that the proposed adaptive testing procedures perform exceptionally well for both dense and sparse alternatives. Motivated by this, we also propose adaptive tests for the sphericity testing problem in high-dimensional settings. First, we establish that the sum-type test $T_{NS}$ is asymptotically independent of the max-type test $T_{NM}$ under the null hypothesis.

\begin{theorem}\label{th3}
Under the same conditions as Theorem \ref{th1} and $p/n\to c\in (0,\infty)$, we have
\begin{align*}
P\left(T_{NS}\le x,T_{NM}\le y\right)\to \Phi(x)G(y).
\end{align*}
where $\Phi(x)$ represents the cumulative distribution function of the standard normal distribution $N(0,1)$.
\end{theorem}

Based on Theorem \ref{th3}, we combine the corrected John's test $T_{NS}$ with $T_{NM}$ using the Cauchy combination test procedure \citep{liu2020} (abbreviated as CN hereafter), i.e.
\begin{align}
T_{CN}=1-F[0.5\tan\{(0.5-p_{NS})\pi\}I(p_{NS}<0.5)+0.5\tan\{(0.5-p_{NM})\pi I(p_{NM}<0.5)\}]
\end{align}
where
\begin{align*}
p_{NS}=1-\Phi(T_{NS}),
p_{NM}=1-G(T_{NM})
\end{align*}
and $F(x)$ is the cumulative distribution function of standard Cauchy distribution $C(0,1)$.

Next, we analyze the power of the adaptive testing procedure. We consider a special case of alternative hypothesis:
\begin{align}\label{h11}
H_1: \bp\not=\bm 0, \|\bp\|_0=o(p^2/\log p), \tr(\bp^2)=O(p/n)
\end{align}
where the diagonal elements of $\bp$ are all equal to zero.

\begin{theorem}\label{th6}
If $\X_i$ follows multivariate normal distribution $N(\bth,{\bms})$, under the alternative hypothesis (\ref{h11}) and $p=O(n)$, we have
\begin{align*}
P\left(T_{NS}\le x,T_{NM}\le y\right)\to P\left(T_{NS}\le x\right)P\left(T_{NM}\le y\right).
\end{align*}
\end{theorem}

Theorem \ref{th6} demonstrates that $T_{NS}$ remains asymptotically independent of $T_{NM}$ under the specific alternative hypothesis (\ref{h11}). According to \cite{li2023}, the power of the test based on the Cauchy combination is superior to that of the test based on the minimum. Furthermore, by applying the principle of inclusion-exclusion, we obtain:
\begin{align}\label{power_H1np}
\beta_{CN, \alpha} \geq \max\{\beta_{{NM}, \alpha/2}, \beta_{{NS}, \alpha/2}, \beta_{{NM}, \alpha/2} + \beta_{{NS}, \alpha/2} - \beta_{{NM}, \alpha/2}\beta_{{NS}, \alpha/2}\}
\end{align}
where $\beta_{CN, \alpha}$, $\beta_{NS, \alpha}$, and $\beta_{NM, \alpha}$ represent the power functions of the CN, NS, and NM tests, respectively.

Next, we also demonstrate that the spatial-sign based sum-type test $T_{SS}$ is asymptotically independent of the spatial-sign based max-type test $T_{SM}$.

\begin{theorem}\label{th4}
Under the same conditions as Theorem \ref{th2} and $p=O(n^2)$, we have
\begin{align*}
P\left(T_{SS}\le x,T_{SM}\le y\right)\to \Phi(x)G(y)
\end{align*}
\end{theorem}

Consequently, we also proposed the following Cauchy combination test (abbreviated as CS hereafter):
\begin{align}
T_{CS}=1-F[0.5\tan\{(0.5-p_{SS})\pi\}I(p_{SS}<0.5)+0.5\tan\{(0.5-p_{SM})\pi I(p_{SM}<0.5)\}]
\end{align}
where
\begin{align*}
p_{SS}=1-\Phi(T_{SS}),
p_{SM}=1-G(T_{SM}).
\end{align*}

For power analysis, we consider the same special alternative hypothesis as (\ref{h11}).

\begin{theorem}\label{th5}
If $\X_i$ follows an elliptical distribution $E_p(\bth,{\bf\Theta})$, under the alternative hypothesis (\ref{h11}) and $p=O(n^2)$, we have
\begin{align*}
P\left(T_{SS}\le x,T_{SM}\le y\right)\to P\left(T_{SS}\le x\right)P\left(T_{SM}\le y\right).
\end{align*}
\end{theorem}

So we also prove that under the special alternative hypothesis (\ref{h11}), the asymptotical independence between $T_{SS}$ and $T_{SM}$ still hold. Then, we also have the following inequality of the power functions:
\begin{align}\label{power2}
\beta_{CS, \alpha} \geq \max\{\beta_{{SM}, \alpha/2}, \beta_{{SS}, \alpha/2}, \beta_{{SM}, \alpha/2} + \beta_{{SS}, \alpha/2} - \beta_{{SM}, \alpha/2}\beta_{{SS}, \alpha/2}\}
\end{align}
where $\beta_{CS, \alpha}$, $\beta_{SS, \alpha}$, and $\beta_{SM, \alpha}$ represent the power functions of the CS, SS, and SM tests, respectively.

\section{Simulation}
The errors are generated from three scenarios:
\begin{itemize}
\item[(I)] Multivariate normal distribution. $\X_i\sim N(\bm\theta,\bms)$.
\item[(II)] Multivariate $t$-distribution $t_{N,3}/\sqrt{3}$.   $\X_i$'s are generated from standardized $t_{N,3}/\sqrt{3}$ with mean zero and scatter matrix $\bms$.
\item[(III)] Multivariate mixture normal distribution $\mbox{MN}_{N,\kappa,9}$. $\X_i$'s are generated from standardized  $[\kappa
N(\bm 0,\bms)+(1-\kappa)N(\bm 0,9\bms)]/\sqrt{\kappa+9(1-\kappa)}$, denoted
by $\mbox{MN}_{N,\gamma,9}$. $\kappa$ is chosen to be 0.8.
\item[(IV)] Independent Component Model. $\X_i=\bms^{1/2}\Z_i$ where $\Z_i=(Z_{i1},\cdots,Z_{ip})$ and $Z_{ij}$s are all i.i.d from $Gamma(4,2)-2$.
\end{itemize}
We examined sample sizes of $n=300$ and $n=600$ with dimensions $p=100, 200, 300, 400$. Table \ref{tab1} displays the results of these six tests under the null hypothesis. Our observations indicate that the three tests based on the spatial-sign method, namely $T_{SS}$, $T_{SM}$, and $T_{CS}$, effectively maintain empirical sizes under elliptical distributions (Scenarios I-III). Conversely, the three tests utilizing the sample covariance matrix, $T_{NS}$, $T_{NM}$, and $T_{CN}$, demonstrate significant size distortion in heavy-tailed distributions, particularly in Scenarios II and III. In the context of the independent component model, the spatial-sign based max-type test $T_{SM}$ exhibits notable size distortion. This outcome is understandable as the independent component model does not adhere to the elliptical distribution assumption when $Z_{ij}$ is non-normal. However, the remaining tests maintain appropriate empirical sizes under Scenario IV.


\begin{table}[!ht]
\begin{center}
\caption{\label{tab1} Sizes of tests.}
                     \renewcommand{\arraystretch}{0.7}
                     \setlength{\tabcolsep}{5pt}{
\begin{tabular}{c|cccc|cccc}
\hline \hline
 Methods & \multicolumn{4}{c}{{$n=300$}} & \multicolumn{4}{c}{{$n=600$}}\\ \hline
$p$& 100&200&300&400& 100&200&300&400\\ \hline
&\multicolumn{8}{c}{Multivariate Normal Distribution}\\ \hline
$T_{SS}$&5.9&5.4&4.5&5.5&4.8&4.4&5&4.4\\
$T_{SM}$&6.5&7.1&6.8&6.5&6.5&7&4.5&5.3\\
$T_{CS}$&6.5&6.9&5.8&7.5&5.9&5.4&5.3&4.1\\
$T_{NS}$&6.3&4.9&4.8&5.5&5.1&4.6&5.1&4.6\\
$T_{NM}$&4.2&3.6&2.9&2.7&5&3.8&3.7&3.1\\
$T_{CN}$&6&4.6&4&4.7&4.5&4.1&5.5&3.2\\ \hline
&\multicolumn{8}{c}{Multivariate $t$ Distribution}\\ \hline
$T_{SS}$&5.6&4.9&5.8&4.7&5.5&4.9&5.5&7\\
$T_{SM}$&4.2&6.8&7.6&7.6&5.2&5.2&5.6&5.6\\
$T_{CS}$&4.8&6.3&6.4&6.1&5&6.1&6.3&5.9\\
$T_{NS}$&100&100&100&100&100&100&100&100\\
$T_{NM}$&100&100&100&100&100&100&100&100\\
$T_{CN}$&100&100&100&100&100&100&100&100\\ \hline
&\multicolumn{8}{c}{Multivariate Mixture Normal Distribution}\\ \hline
$T_{SS}$&4.2&5.2&4&5.5&4.6&4.3&5.6&6.1\\
$T_{SM}$&5.6&6.3&7.2&7.5&5.8&5.1&6.8&5.2\\
$T_{CS}$&5.4&5.4&5.4&7.4&5&4.9&6.2&6\\
$T_{NS}$&100&100&100&100&100&100&100&100\\
$T_{NM}$&100&100&100&100&100&100&100&100\\
$T_{CN}$&100&100&100&100&100&100&100&100\\ \hline
&\multicolumn{8}{c}{Independent Component Model}\\ \hline
$T_{SS}$&3.9&3.5&6.1&4.9&4.7&5.6&5.3&4.9\\
$T_{SM}$&15.5&22.6&27.8&30.3&11.9&16.3&19&19.6\\
$T_{CS}$&12.3&17&21.6&24.2&9.4&13.2&14.6&15.1\\
$T_{NS}$&4.8&4.3&7.4&5.2&5.8&6.8&5.7&5.6\\
$T_{NM}$&7.8&7.1&8.4&7.2&5.4&5.3&6.9&5.6\\
$T_{CN}$&7.5&6.1&8.6&8.2&5.8&6.9&6.6&5.6\\
\hline
\hline
\end{tabular}}
\end{center}
\end{table}

\begin{figure}[!ht]
\caption{Power of tests with different sparsity levels with $(n,p)=(100,200)$. \label{power}}
\centering
{\includegraphics[width=1\textwidth]{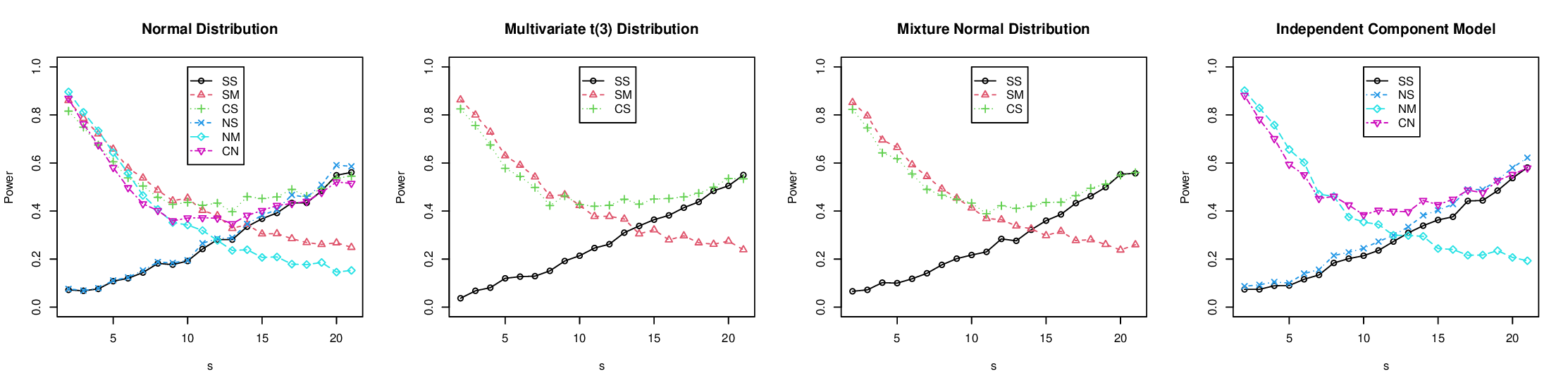}}
\end{figure}

To assess the power performance of each test, we adopt the alternative hypothesis characterized by $\bms = \text{diag}\{{\bf A}_s, \I_{p-s}\}$, where ${\bf A}_s = (1-\delta)\I_s + \delta\bm 1\bm 1^T$ and $\delta = 0.5s^{-1/2}$. Given that the empirical sizes of $T_{NS}$, $T_{NM}$, and $T_{CN}$ are considerably large, we exclude these three tests from Scenarios II and III. And we also only consider $T_{NS}, T_{NM}, T_{CN}$ and $T_{SS}$ for Scenario IV because the sizes of $T_{SM}$ are larger than the nominal level in this case. Figure \ref{power} illustrates the power curves for each test across different sparsity levels $s$ with sample dimensions $(n, p) = (300, 400)$.

First, we consider the multivariate normal distribution. For sparse alternatives, where the sparsity level $s$ is small ($s < 10$), the sum-type tests ($T_{NS}$ and $T_{SS}$) exhibit lower power compared to the max-type tests ($T_{NM}$ and $T_{SM}$). Conversely, for dense alternatives, the sum-type tests ($T_{NS}$ and $T_{SS}$) demonstrate better performance than the max-type tests ($T_{NM}$ and $T_{SM}$). The Cauchy combination tests ($T_{CN}$ and $T_{CS}$) exhibit power performance comparable to the best of both sum-type and max-type tests. When the sparsity level is moderate, the $T_{CN}$ and $T_{CS}$ tests achieve the highest power, highlighting the advantages of adaptive-type test procedures. For heavy-tailed distributions (Scenario II and III), the $T_{CS}$ test consistently demonstrates nearly optimal performance across all scenarios. For the independent component model, $T_{SS}$ has similar performance as $T_{NS}$. The adaptive method $T_{CN}$ also nearly has the best performance in all cases.

\begin{figure}[htbp]
\centering
\caption{Power of tests with different signal strength. \label{figs}}
\subfloat[$s=2$]{\includegraphics[width=1\textwidth]{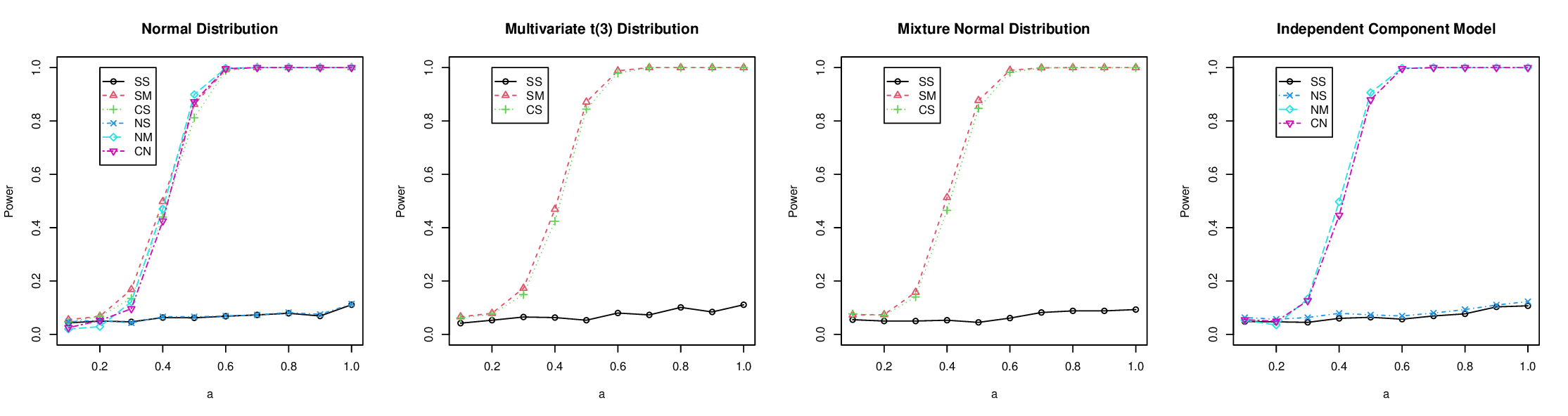}}\\
\subfloat[$s=10$]{\includegraphics[width=1\textwidth]{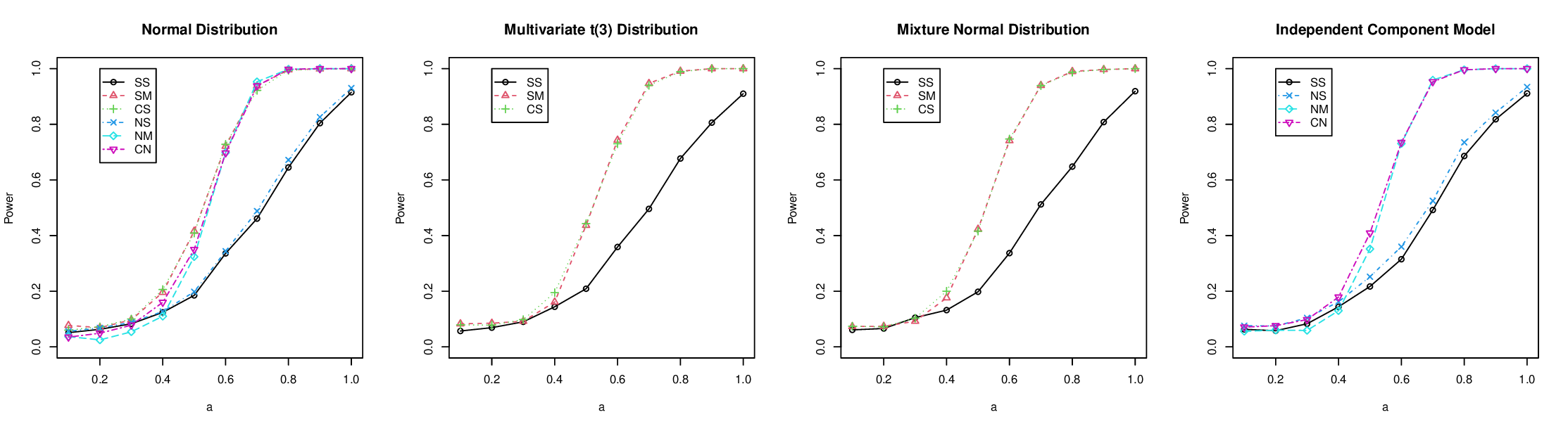}}\\
\subfloat[$s=30$]{\includegraphics[width=1\textwidth]{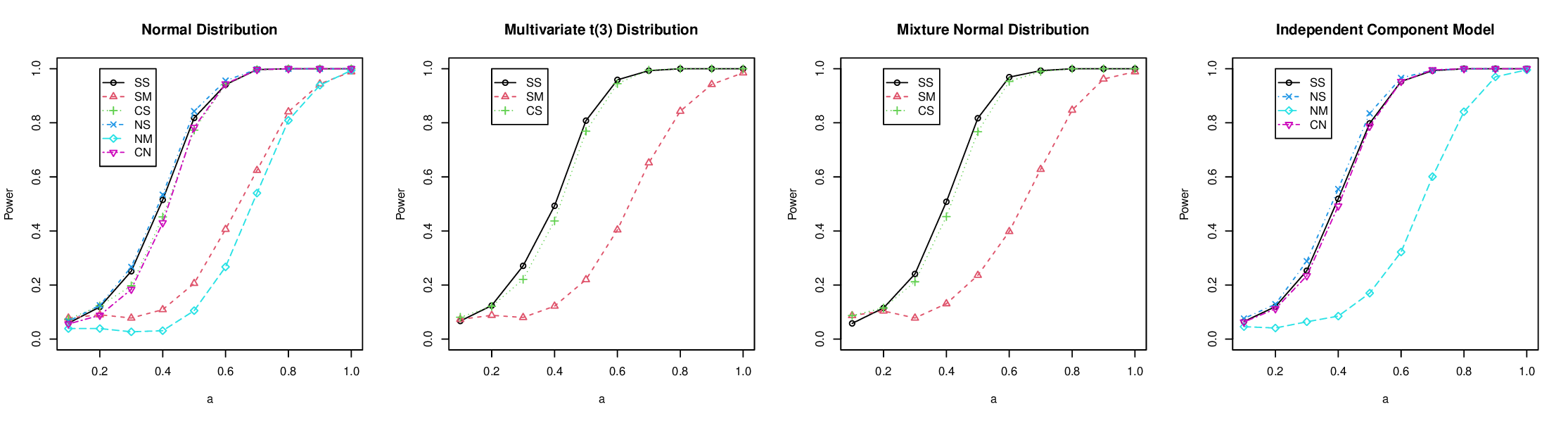}}
\end{figure}

To assess the performance of these tests under varying signal strengths, we consider the alternative hypothesis defined by $\delta = as^{-1/2}$ with three different sparsity levels: $s = 2, 10, 30$. All other experimental settings remain consistent with previous analyses. Figure \ref{figs} presents the corresponding power curves for each test across different signal strengths. Our findings indicate that the power of all tests increases as the signal strength increases, demonstrating the consistency of each test. Similar to the results shown in Figure \ref{power}, sum-type tests exhibit lower power compared to max-type tests under sparse alternatives, whereas the opposite trend is observed for dense alternatives. Additionally, the adaptive tests consistently demonstrate superior performance, often achieving the best results overall. These simulation results collectively underscore the advantages of our proposed adaptive procedures.

\section{Conclusion}
To address the challenge of sparse alternative hypotheses in sphericity tests within high-dimensional contexts, we introduced two distinct max-type test procedures. One is based on the assumption of an independent component model, while the other relies on an elliptical distribution assumption. Additionally, we proposed two Cauchy combination test procedures capable of managing both dense and sparse alternatives. In practical applications, the choice of test procedure can be tailored to the specific distribution assumption. Furthermore, spatial-rank has proven to be a highly effective method for sphericity testing in high dimensions, as evidenced in \cite{feng2017}. Consequently, developing a max-type test procedure based on spatial-rank for high-dimensional sphericity tests, along with a corresponding adaptive test procedure, merits further exploration and research.

\section{Appendix}
\subsection{Proof of Theorem \ref{th1}}
By the Bernstein inequality, we can easily obtain that $\max_{1\le k\le p}(\hat{\kappa}_k-\kappa_k)=O_p(\sqrt{\log p/n})$ and $\max_{1\le k\le p}(\hat{\sigma}_k^2-\sigma_k^2)=O_p(\sqrt{\log p/n})$. Thus,
\begin{align*}
T_{NM}=&\max\left\{\max_{1\le k\le p}\frac{n(\hat{\sigma}_k^2-p^{-1}\tr(\S))^2}{{\kappa}_k},\max_{1\le i<j\le p}\frac{n\hat\sigma_{ij}^2}{{\sigma}_i^2{\sigma}_j^2}\right\}\\
&-2\log(p(p+1)/2)+\log\log(p(p+1)/2)+O_p(n^{-1/2}\log^{3/2} p).
\end{align*}
Additionally, under the null hypothesis, $p^{-1}\tr(\S)=\sigma^2+O_p((np)^{-1/2})$. So
\begin{align*}
T_{NM}=&\max\left\{\max_{1\le k\le p}\frac{n(\hat{\sigma}_k^2-\sigma^2)^2}{{\kappa}_k},\max_{1\le i<j\le p}\frac{n\hat\sigma_{ij}^2}{{\sigma}_i^2{\sigma}_j^2}\right\}\\
&-2\log(p(p+1)/2)+\log\log(p(p+1)/2)+O_p(n^{-1/2}\log^{3/2} p)+O_p(p^{-1/2}\log^{1/2} p).
\end{align*}
Define $w_l=n^{-1/2}\sum_{i=1}^n\sigma\kappa_l^{-1/2}(z_{il}^2-1)$ for $l=1,\cdots,p$, $w_l=n^{-1/2}\sum_{k=1}^nz_{ki}z_{kj}, l=\sum_{m=0}^{i-1}(n-m-1)+(j-i-1)+p$ and $\bm w=(w_1,\cdots,w_{p(p+1)/2})^T$. Under the independence assumption of $z_{ij}$, we have $\cov(\bm w)=\I_{p(p+1)/2}$. By Theorem 2.1 in \cite{chernozhuokov2022}, we have
\begin{align*}
\sup_{A\in \mathcal{A}}\left|P\left(\bm w\in A\right)-P\left(\bm Y \in A\right)\right|\le C\left(\frac{\log^5(p^2n)}{n}\right)^{1/4}
\end{align*}
where $\bm Y=(Y_1,\cdots,Y_{p(p+1)/2})^T \sim N\left(0, \I_{p(p+1)/2}\right)$ and $\mathcal{A}$ is the class of all hyper-rectangles in $\mathbb{R}^{p(p+1)/2}$, i.e. sets of the form
$$
A=\left\{w=\left(w_1, \ldots, w_{p(p+1)/2}\right)^{\prime} \in \mathbb{R}^p: a_{l j} \leq w_j \leq a_{r j} \text { for all } j=1, \ldots, {p(p+1)/2}\right\},
$$
for some constants $-\infty \leq a_{l j} \leq a_{r j} \leq \infty$ with $j=1, \ldots, {p(p+1)/2}$. Thus,
\begin{align*}
P\left(\max_{1\le k\le p(p+1)/2} w_i^2 \le x\right)-P\left(\max_{1\le k\le p(p+1)/2} Y_i^2 \le x\right)\to 0
\end{align*}
by the assumption $\frac{\log^5(p^2n)}{n}\to 0$. By Lemma 6 in \cite{Cai2014}, we have
\begin{align*}
P\left(\max_{1\le k\le p(p+1)/2} Y_i^2-2\log(p(p+1)/2)+\log\log(p(p+1)/2) \le x\right)\to G(x).
\end{align*}
Here we obtain the result. \hfill$\Box$

\subsection{Proof of Theorem \ref{th2}}
Define $\tilde{\O}=\frac{1}{n}\sum_{i=1}^n U(\X_i-{\bth})U(\X_i-{\bth})^T\doteq (\tilde \psi_{ij})_{1\le i,j\le p}$. First, we show that
\begin{align*}
\tilde T_{SM}=&\max\left\{\max_{1\le i\le p}{np(p+2)\frac{(\tilde\psi_{ii}-p^{-1})^2}{2(1-p^{-1})}},\max_{1\le i<j\le p}np(p+2)\tilde\psi_{ij}^2\right\}\\
&-2\log(p(p+1)/2)+\log\log(p(p+1)/2)
\end{align*}
has the limit null distribution
\begin{align}\label{tsm1}
P\left(\tilde T_{SM}\le x\right)\to G(x).
\end{align}
Define $\mathbf{K}_{p, p}$ is the commutation matrix, that is, a $p^2 \times p^2$ block matrix with $(i, j)$-block being equal to a $p \times p$ matrix that has one at entry $(j, i)$ and zero elsewhere, and $\mathbf{J}_{p, p}$ for $\operatorname{vec}\left(\mathbf{I}_p\right) \operatorname{vec}\left(\mathbf{I}_p\right)^{\prime}$.
Define
$$
\mathbf{C}_{p, p}=\frac{1}{2}\left(\mathbf{I}_{p^2}+\mathbf{K}_{p, p}\right)-\frac{1}{p} \mathbf{J}_{p, p}
$$
projects a vectorized matrix vec $(\mathbf{A})$ to the space of symmetrical and centered vectorized matrices. Define $\vech(\A)=\M_p\vec(\A)$ where $\vech(\A)$ is the half-vectorization of a symmetric matrix $\A\in \mathbb{R}^p$, which includes elements on or above the diagonal and $\M_p$ is a transformation matrix that maps $\vec(\A)$ to $\vech(\A)$. Under the null hypothesis,
$$
\mathbf{E}(\operatorname{vech}(\tilde{\O}))=\frac{1}{p} \operatorname{vech}\left(\mathbf{I}_p\right) \quad \text { and } \quad \cov(\operatorname{vech}(\tilde{\O}))=\frac{2}{np(p+2)} \M_{p^2}\mathbf{C}_{p, p}\M_{p^2}^T,
$$
and therefore also
$$
\mathbf{E}\left(\M_{p^2}\mathbf{C}_{p, p} \operatorname{vec}(\tilde{\O})\right)=\mathbf{0} \quad \text { and } \quad \cov\left(\M_{p^2}\mathbf{C}_{p, p} \operatorname{vec}(\tilde{\O})\right)=\frac{2}{np(p+2)}\M_{p^2} \mathbf{C}_{p, p}\M_{p^2}^T.
$$
Let $\bm u=\B^{-1/2}\M_{p^2}\mathbf{C}_{p, p} \operatorname{vec}(\tilde{\O})$ where $\B$ is the diagonal matrix of $\frac{2}{np(p+2)} \M_{p^2}\mathbf{C}_{p, p}\M_{p^2}^T$. So
\begin{align*}
\|\bm u\|_\infty=\max\left\{\max_{1\le i\le p}{np(p+2)\frac{(\tilde\psi_{ii}-p^{-1})^2}{2(1-p^{-1})}},\max_{1\le i<j\le p}np(p+2)\tilde\psi_{ij}^2\right\}
\end{align*}
By Theorem 2.1 in \cite{chernozhuokov2022}, we have
\begin{align*}
\sup_{A\in \mathcal{A}}\left|P\left(\bm u\in A\right)-P\left(\bm Y \in A\right)\right|\le C\left(\frac{\log^5(p^2n)}{n}\right)^{1/4}
\end{align*}
where $\bm Y=(Y_1,\cdots,Y_{p(p+1)/2})^T \sim N\left(0, {\bf \Xi}\right)$ where ${\bf \Xi}=\frac{2}{np(p+2)} \B^{-1/2}\M_{p^2}\mathbf{C}_{p, p}\M_{p^2}^T\B^{-1/2}$ and $\mathcal{A}$ is the class of all hyper-rectangles in $\mathbb{R}^{p(p+1)/2}$, i.e. sets of the form
$$
A=\left\{w=\left(w_1, \ldots, w_{p(p+1)/2}\right)^{\prime} \in \mathbb{R}^{p(p+1)/2}: a_{l j} \leq w_j \leq a_{r j} \text { for all } j=1, \ldots, p(p+1)/2\right\},
$$
for some constants $-\infty \leq a_{l j} \leq a_{r j} \leq \infty$ with $j=1, \ldots, p(p+1)/2$.
Thus,
\begin{align*}
P\left(\|\bm u\|_\infty^2 \le x\right)-P\left(\|\bm Y\|_\infty^2 \le x\right)\to 0
\end{align*}
by the assumption $\frac{\log^5(p^2n)}{n}\to 0$. By Lemma 6 in \cite{Cai2014}, we have
\begin{align*}
P\left(\|\bm Y\|_\infty^2-2\log(p(p+1)/2)+\log\log(p(p+1)/2) \le x\right)\to G(x).
\end{align*}
So we obtain the result (\ref{tsm1}). Similarly, define $\hat{\bm u}=\B^{-1/2}\mathbf{C}_{p, p} \operatorname{vec}(\hat{\O})$. Then,
$$T_{SM}-\tilde{T}_{SM}=\|\hat{\bm u}\|_\infty^2-\|{\bm u}\|_\infty^2.$$
By the definition, we have
\begin{align*}
\|\hat{\bm u}\|_\infty-\|{\bm u}\|_\infty=O_p(\sqrt{np^2})\|\tilde{\O}-\hat\O\|_\infty=O_p(\|\hat{\bth}-\bth\|_\infty).
\end{align*}
According to Corollary 1 in \cite{cheng2023}, we have $\|\hat{\bth}-\bth\|_\infty=O_p(\sqrt{\log p/n})$. Additionally, by (\ref{tsm1}), $\|\hat{\bm u}\|_\infty+\|{\bm u}\|_\infty=O_p(\log^{1/2} p)$. Then, $T_{SM}-\tilde{T}_{SM}=O_p(n^{-1/2}\log p)$ by triangle inequality. So, by the assumption of $p$ and $n$, we can easily obtain the result. \hfill$\Box$

\subsection{Proof of Theorem \ref{th3}}
According to the proof of Theorem 2.2 in \cite{Wang2013}, we have
\begin{align*}
T_{NS}=\frac{\sum_{i=1}^{p(p+1)/2}w_i^2-p(p+1)/2}{\sqrt{p(p+1)}}+o_p(1).
\end{align*}
And by the proof of Theorem \ref{th1}, we have
\begin{align*}
T_{NM}=\|\bm w\|_\infty^2-2\log(p(p+1)/2)+\log\log(p(p+1)/2)+o_p(1).
\end{align*}
Taking the same procedure as the proof of Theorem 6 in \cite{liu2024}, we can obtain that
\begin{align*}
&P\left(\frac{\bm w^T \bm w-p(p+1)/2}{\sqrt{p(p+1)}}\le x,\|\bm w\|_\infty^2-2\log(p(p+1)/2)+\log\log(p(p+1)/2)\le y\right)\\
=&P\left(\frac{\bm Y^T\bm Y-p(p+1)/2}{\sqrt{p(p+1)}}\le x,\|\bm Y\|_\infty^2-2\log(p(p+1)/2)+\log\log(p(p+1)/2)\le y\right)+o(1)
\end{align*}
where $\bm Y\sim N(\bm 0, \I_{p(p+1)/2})$. According to Theorem 3 in \cite{Feng2024}, we have
\begin{align*}
&P\left(\frac{\bm Y^T\bm Y-p(p+1)/2}{\sqrt{p(p+1)}}\le x,\|\bm Y\|_\infty^2-2\log(p(p+1)/2)+\log\log(p(p+1)/2)\le y\right)\\
&\to P\left(\frac{\bm Y^T\bm Y-p(p+1)/2}{\sqrt{p(p+1)}}\le x\right)P\left(\|\bm Y\|_\infty^2-2\log(p(p+1)/2)+\log\log(p(p+1)/2)\le y\right)\\
&=\Phi(x)G(y).
\end{align*}
Finally, by Lemma S10 in \cite{Feng2024}, we obtain the result
\begin{align*}
P\left(T_{NS}\le x,T_{NM}\le y\right)\to \Phi(x)G(y).
\end{align*}
 \hfill$\Box$

\subsection{Proof of Theorem \ref{th6}}
Without loss of generality, we assume that $\sigma^2=1$.
Under the assumption of normality and the diagonal matrix of $\G$ are all equal to zero, we can easily obtain that
\begin{align*}
T_{NS}=&\frac{n}{2p}(\tr(\S^2)-p)-\frac{p+1}{2}+o_p(1)\\
=&\frac{n}{2p}\sum_{1\le i,j\le p}\hat{\sigma}_{ij}^2-\frac{n+p+1}{2}+o_p(1).
\end{align*}
Accordingly, we also have
\begin{align*}
T_{NM}=&\max\left\{\max_{1\le k\le p}\frac{n(\hat{\sigma}_k^2-1)^2}{{\kappa}_k},\max_{1\le i<j\le p}{n\hat\sigma_{ij}^2}\right\}\\
&-2\log(p(p+1)/2)+\log\log(p(p+1)/2)+o_p(1).
\end{align*}
Then, taking the same procedure as the proof of the following Theorem \ref{th5}, we can also obtain the result. \hfill$\Box$

\subsection{Proof of Theorem \ref{th4}}
According to the proof of Theorem 1 in \cite{Zou2014}, we have
\begin{align*}
T_{SS}=\frac{\bm u^T\bm u-p(p+1)/2}{\sqrt{p(p+1)}}+o_p(1)
\end{align*}
And by the proof of Theorem \ref{th2}, we have
\begin{align*}
T_{SM}=\|\bm u\|_\infty^2-2\log(p(p+1)/2)+\log\log(p(p+1)/2)+o_p(1).
\end{align*}
Then, taking the same procedure as the proof of Theorem \ref{th3}, we also can obtain that
\begin{align*}
P\left(T_{SS}\le x,T_{SM}\le y\right)\to \Phi(x)G(y).
\end{align*}
 \hfill$\Box$

\subsection{Proof of Theorem \ref{th5}}
According to the proof of Theorem 2 in \cite{Zou2014}, we have
\begin{align*}
T_{SS}=&\frac{\bm u^T\bm u-p(p+1)/2}{\sqrt{p(p+1)}}+np^{-1}\tr(\bp^2)+o_p(1)\\
=&\frac{\sum_{i\in \mathcal{M}}u_i^2+\sum_{i\in \mathcal{M}^c}u_i^2-p(p+1)/2}{\sqrt{p(p+1)}}+np^{-1}\tr(\bp^2)+o_p(1)
\end{align*}
where $\mathcal{M}$ is the nonzero elements of $\vech(\bp)$.
Obviously, we can show that $u^2_i-1$'s are independent sub-exponential random variables. So
\begin{align*}
P\left(p^{-1}\sum_{i\in \mathcal{M}}(u_i^2-1)\ge \epsilon\right)\le K_\epsilon^m e^{-c_\epsilon p^{1/2}}\to 0,
\end{align*}
where $K_\epsilon$ and $c_\epsilon$ are two positive constants dependent on $\epsilon$. So
\begin{align*}
T_{SS}=\frac{\sum_{i\in \mathcal{M}^c}u_i^2-|\mathcal{M}^c|}{\sqrt{p(p+1)}}+np^{-1}\tr(\bp^2)+o_p(1).
\end{align*}
Taking the same procedure as the proof of Theorem~\ref{th2}, we have $T_{SM}=\breve{T}_{SM}+o_p(1)$, where
\begin{align*}
\breve{T}_{SM}=\max _{1 \leq i \leq p(p+1)/2}\left(u_i+\tilde{\mu}_i\right)^2=\max \left\{\max _{i \in \mathcal{M}}\left(u_i+\tilde{\mu}_i\right)^2, \max_{i \in \mathcal{M}^c} u_i^2\right\}.
\end{align*}
Taking the same procedure as the proof of Theorem 3 in \cite{Feng2024}, by the weak correlation of $u_i$, $\sum_{i\in \mathcal{M}^c}u_i^2$ is asymptotically independent of $\max _{i \in \mathcal{M}}\left(u_i+\tilde{\mu}_i\right)^2$. According to Theorem~\ref{th4}, $\sum_{i\in \mathcal{M}^c}u_i^2$ is asymptotically independent of $\max_{i \in \mathcal{M}^c} u_i^2$. So we obtain that $T_{SS}$ is asymptotically independent of $T_{SM}$. \hfill$\Box$

\end{document}